\documentclass{kapproc} 

\usepackage{procps} 

\usepackage[dvips]{graphicx}

\upperandlowercase

\kluwerbib 

\begin{document}

\articletitle{The distribution of kHz QPO frequencies in Sco X-1}


\author{Tomaso Belloni}
\affil{INAF -- Osservatorio Astronomico di Brera, Merate, Italy}
\email{belloni@merate.mi.astro.it}

\author{Mariano Mendez}
\affil{SRON, Utrecht, the Netherlands}
\email{M.Mendez@sron.nl}

\author{Jeroen Homan}
\affil{MIT, Cambridge, USA}
\email{jeroen@space.mit.edu}

\begin{abstract}
The frequencies of kHz Quasi-Periodic Oscillations in the bright Low-Mass
X-ray Binary Sco X-1 have been reported to appear preferentially in a 3:2
ratio. We show that a proper statistical analysis of those RXTE data
upon which the claim is based indicated a 2.4$\sigma$ deviation from a
constant distribution in ratios. The analysis of a much larger RXTE/PCA dataset
shows that there is no sharp concentration around a  3:2 ratio, but that the
ratios are broadly distributed over the range 820--1150 Hz.
\end{abstract}


\section*{Introduction}
Recently, Abramowicz et al. (2003, hereafter A03) 
reported that the ratio of the frequencies in the
kHz Quasi-Periodic Oscillations (QPO) from the brightest 
LMXB in the sky, Sco X-1, tend to cluster around a value of 1.5, which they
interpret as evidence for a resonance.
A mathematical approach to the resonance model was presented by Rebusco
(2004) in order to explain the Sco X-1 results; the discrepancies of the data
with a pure 3:2 ratio were attributed to the action of an additional
ad-hoc force. 
Here, we present a statistically correct method to establish the presence
of such a ratio and analyze a larger RXTE dataset from Sco X-1.

\section{The analysis of kHz QPO ratios}

It is known that the kHz QPOs in bright LMXBs change frequencies while
keeping their difference approximately constant (see van der Klis 2004 for
a review). This implies a roughly linear relation between them, by 
definition not compatible with a fixed ratio. This can be seen in Fig. 1,
where the values for Sco X-1 used by A03 are plotted together with a sample
of published values for both atoll and Z sources.

\begin{figure}[ht]
\includegraphics*[width=12.0 true cm]{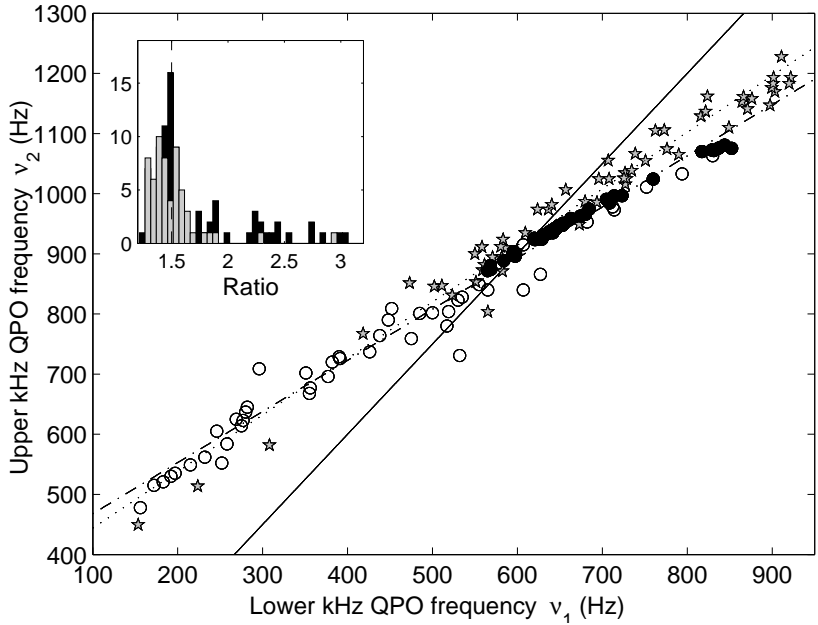}
\caption{Correlation between lower and upper kHz QPO frequency for the
	Sco X-1 from A03 [black circles], a sample
	of atoll sources [open circles] and Z sources [stars]
	from the literature. The line represents a fixed 3:2 ratio.
	In the inset, the distributions of ratios for the atoll (black) and
	Z sources (gray) are shown.
        }
\end{figure}

The inset in Fig. 1 shows the corresponding distributions in $\nu_2 / \nu_1$
for atoll and Z sources: the peaks at $\sim$1.5 do not correspond to an
over-density of points in the $\nu_2$ vs. $\nu_1$ plot. The reason is that
the ratio of two variables linearly correlated ($y=ax+b$) with $b\neq0$ is
determined mostly by the distribution in one of the variables. 
Even a homogeneous distribution of points along the correlation, within
certain bounds, would lead to a peaked distribution of ratios.
This means
that determining the distribution of ratios in this case is {\it not} 
the correct
 statistical way to assess whether there is an excess around that ratio.

A correct statistical procedure consists in measuring the distribution of
one of the two frequencies ($\nu_2$ in Fig. 2, black histogram) with 
respect to the value corresponding to a 3:2 ratio, as estimated from a 
linear fit to the points in Fig. 1. This fit for the Sco X-1 data yields
$a=0.768\pm 0.02$ and $b=432.5\pm 1.5$ and corresponds to the vertical dashed
line in Fig. 2. The black histogram in Fig. 2 shows a peak around the
expected 3:2 frequency, but it is inconsistent with a constant
distribution only at a 2.4$\sigma$ level, due to the low statistics.

\section{The large sample}

We accumulated a much larger sample of $\nu_2$ for Sco X-1 using the data
analyzed by Mendez \& van der Klis (2000). Our sample consists of 1393
frequencies, accumulated from RXTE/PCA data stretches 128s long (see
Mendez \& van der Klis 2000 for details about the extraction). The
distribution of these points is shown in Fig. 2 (gray histogram).
This distribution shows three (possibly random) peaks which we fitted with
Gaussians. The broadest one is centered at $\nu_2$=936$\pm$2 Hz,
inconsistent with the value of 
 $\nu_2$=909$\pm$2 Hz, corresponding to a 3:2
ratio along the correlation of Fig. 1.
Notice that, as the upper kHz QPO is much easier to detect than the lower
kHz QPO in Sco X-1, we do not have a corresponding distribution in $\nu_1$.
Although we have shown that the other frequency is not necessary to determine
the ratio, once the correlation in Fig. 1 is established, this procedure
relies on the assumption of the {\it existence} of the lower peak and on
its frequency following that relation. However, it is currently the best
that can be done with the existing instruments.
A full analysis of this distribution and those from other bright LMXBs
will be presented in a forthcoming paper.

\begin{figure}[ht]
\includegraphics*[width=12.0 true cm]{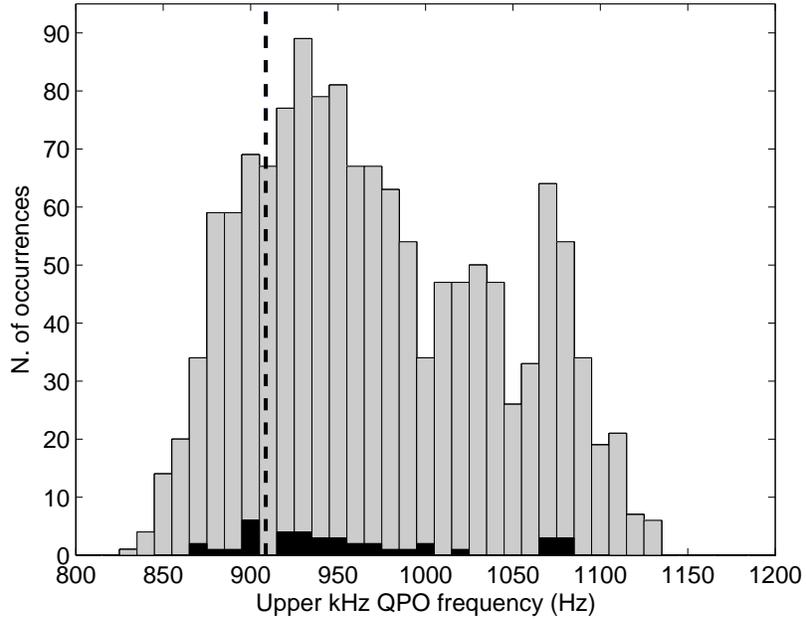}
\caption{Distribution of the occurrence of the upper kHz QPO
	frequency for Sco X-1 from our sample. The black histogram 
	is the the data from A03. The 
	dashed line indicates the frequency corresponding to a 3:2
	ratio with the lower kHz QPO frequency (see text). }
\end{figure}

\section{Conclusions}

We have shown that a correct statistical treatment of a large set of data
allows us to examine the distribution in kHz QPO frequencies for Sco X-1:
in this broad distribution, the values corresponding to a 3:2 ratio do not
appear to be preferred. 
It is known that the fractional rms  of the kHz
QPO peaks decreases at high and at low frequencies (see Mendez, van der Klis 
\& Ford 2001; Di Salvo, Mendez \& van der Klis 2003), which is what is
observed here. Our results indicate that there is not evidence from the data
of a simple 3:2 resonance, which can be excluded at high significance. 
The introduction of an additional force has been shown to be able to move
the frequencies away from the 3:2 ratio (Rebusco 2004): however, this model
has two major disadvantages: the first is that it introduces an unknown 
force, and therefore parameter, in the model; the second is that in moving
away the frequencies from their resonant values, it also removes the only
observational prediction made by the model.

\end{document}